\documentclass[namedreferences]{SolarPhysics}
\usepackage[optionalrh,natbib]{spr-sola-addons} % For Solar Physics 
\usepackage{graphicx}                    % For eps figures, newer & more powerfull
\usepackage{color}                       % For color text: \color command
\usepackage{url}                         % For breaking URLs easily trough lines
\begin{document}

\begin{article}

\begin{opening}

\title{Solar Cycle Variation of Magnetic Flux Ropes in a Quasi-Static Coronal Evolution Model}

%%%%%%%%%%%%%%%%%%%%%%%%%%%%%%%%%%%%%%%%%%%%%%%%%%%
%% Authors Names
%
\author{A.R.~\surname{Yeates}$^{1,4}$\sep
        J.A.~\surname{Constable}$^{2}$\sep
        P.C.H.~\surname{Martens}$^{1,3}$      
       }

%%%%%%%%%%%%%%%%%%%%%%%%%%%%%%%%%%%%%%%%%%%%%%%%%%%
%% Runningheads
%
\runningauthor{A.R. Yeates {\it et al.}}
\runningtitle{Solar Cycle Variation of Magnetic Flux Ropes}

%%%%%%%%%%%%%%%%%%%%%%%%%%%%%%%%%%%%%%%%%%%%%%%%%%%
%% Affilations 
%
  \institute{$^{1}$ Harvard-Smithsonian Center for Astrophysics, 60 Garden Street, Cambridge, MA 02138, USA \\ 
             $^{2}$ School of Mathematics and Statistics, University of St Andrews, St Andrews, KY16 9SS, UK\\
             $^{3}$ Department of Physics, Montana State University, Bozeman, MT 59717, USA\\
          $^{4 }$  Now at: Division of Mathematics, University of Dundee, Dundee, DD1 4HN, UK\\
                     email: \url{anthony@maths.dundee.ac.uk} 
%                     email: \url{e.mail-c} \\
             }

%%%%%%%%%%%%%%%%%%%%%%%%%%%%%%%%%%%%%%%%%%%%%%%%%%%
%%% Abstract 
\begin{abstract}
The structure of electric current and magnetic helicity in the solar corona is closely linked to solar activity over the 11-year cycle, yet is poorly understood. As an alternative to traditional current-free ``potential field'' extrapolations, we investigate a model for the global coronal magnetic field which is non-potential and time-dependent, following the build-up and transport of magnetic helicity due to flux emergence and large-scale photospheric motions. This helicity concentrates into twisted magnetic flux ropes, which may lose equilibrium and be ejected. Here, we consider how the magnetic structure predicted by this model---in particular the flux ropes---varies over the solar activity cycle, based on photospheric input data from six periods of cycle 23. The number of flux ropes doubles from minimum to maximum, following the total length of photospheric polarity inversion lines. However, the number of flux rope ejections increases by a factor of eight, following the emergence rate of active regions. This is broadly consistent with the observed cycle modulation of coronal mass ejections, although the actual rate of ejections in the simulation is about a fifth of the rate of observed events. The model predicts that, even at minimum, differential rotation will produce sheared, non-potential, magnetic structure at all latitudes. 
\end{abstract}

\end{opening}
%-------------------------------------------------

%%%%%%%%%%%%%%%%%%%%%%%%%%%%%%%%%%%%%%%%%%%%%%%%%%%
%% Sections
%
\section{Introduction}\label{sec:intro} 

The number of sunspots has long been known to vary with a period of about 11 years. Since the early 20th century it has been recognised that sunspots carry magnetic flux originating from the Sun's interior. Once this magnetic flux emerges into the corona, it dominates both the structure and evolution of the plasma there. This is evidenced by 11-year variations in many phenomena, including solar flares \citep{aschwanden2005}, prominences \citep{dazambuja1955,hansen1975}, X-ray flux \citep{pevtsov2001}, coronal mass ejections \citep[CMEs;][]{webb1994,cremades2007}, and the solar wind \citep{richardson2001}.

Despite the connection to observed phenomena, the coronal magnetic field cannot usually be measured directly due to the tenuous nature of the plasma, so only isolated measurements exist \citep[{\it e.g.},][]{lin2000,casini2003,tomczyk2007}. Usually, it must be extrapolated from the routinely observed magnetic field in the solar photosphere, perhaps using images of high-temperature loops to constrain the field topology. This topology is often found to contain non-zero electric currents \citep[{\it e.g.} in X-ray sigmoids;][]{canfield1999}, and extrapolation of such magnetic fields is not yet robust, even when limited to a single active region \citep{derosa2009}. On the global scale, which concerns us in this paper, it is usual to assume a current-free magnetic field (a potential field). This is uniquely determined in the corona given an observed radial component on the photosphere and vanishing horizontal components at an upper ``source surface'' \citep[usually placed at $r=2.5R_\odot$;][]{altschuler1969}. Even the more realistic models that solve for MHD equilibria in the corona usually assume a potential magnetic field which is then perturbed to be consistent with a given thermodynamic structure and solar wind \citep{riley2006,cohen2007}. Whilst the potential field model has been successful in describing certain large-scale aspects of coronal magnetic structure---such as the locations of coronal holes \citep{wang1996}---it does not include the development of currents and free magnetic energy that are needed to model solar eruptions. In fact, the highly sheared magnetic fields observed in long-lived filament channels all over the Sun \citep{martin1994} show that, even outside active regions, the assumption of a potential field can be inadequate.

As a first step toward understanding the structure of currents and magnetic helicity in the global corona, we have investigated an alternative model for the coronal magnetic field, based also on observational input of the photospheric magnetic field. In this model, the large-scale mean magnetic field in the corona evolves in time through a quasi-static relaxation, in response to flux emergence and to surface motions \citep{vanballegooijen2000,yeates2008,yeates2009b}. The consequence is that, unlike in a sequence of potential field extrapolations, current and magnetic helicity are generated in and transported through the corona. Flux cancellation above photospheric polarity inversion lines (PILs) then leads to the concentration of helicity in either sheared arcades or twisted magnetic flux ropes \citep[through the mechanism described by][]{vanballegooijen1989}. Comparison of the simulated magnetic field direction at these locations with observations of filament chirality has demonstrated that, despite its simplicity, the model is able to reproduce the general structure of the magnetic field at most of these locations \citep{yeates2008}.

In this paper, we use the quasi-static model to simulate the coronal magnetic field during six distinct periods over cycle 23. Our aim is to consider how the magnetic field structure predicted by this model varies over the 11-year solar activity cycle. We focus on a particular aspect of the field structure: the formation and ejection of magnetic flux ropes. These are not only observed in the real corona, but their ejection is suggested to give rise to CMEs \citep[{\it e.g.},][]{gibson2006}. By their very nature, flux ropes cannot be modelled by potential field extrapolations, so their locations and ejections have not been considered in previous models of the global corona. The formation of flux ropes in the present quasi-static model has been described for a simple two-bipole configuration by \citet{mackay2006}, and for the global corona over a particular 5-month period by \citet{yeates2009b}. Here we extend the latter study to simulate different phases of the solar cycle.

\section{Coronal Magnetic Field Simulations}\label{sec:sims}

Our numerical simulations of the global coronal magnetic field are based on the mean-field model of \citet{vanballegooijen2000}, and were described in detail by \citet{yeates2009b}. The key features are as follows:
\begin{enumerate}
\item The large-scale mean-field in the corona evolves through the non-ideal MHD induction equation, with a turbulent diffusivity to parametrize the effect of small-scale fluctuations.
\item We impose an artificial ``magneto-frictional relaxation'' velocity so that the coronal field responds to photospheric driving by evolving through a sequence of force-free equilibria.
\item The photospheric driver is the standard surface flux transport model for the radial magnetic field \cite{sheeley2005}.
\item New active regions are inserted in the form of idealised magnetic bipoles with properties (location, size, tilt, and magnetic flux) based on observations. The 3D bipoles are given a non-zero helicity.
\end{enumerate}

The simulations use a stretched spherical coordinate grid between radii $R_\odot$ and $2.5 R_\odot$ with an effective resolution of $1^\circ$ at the equator \citep[see][]{yeates2008}. As observational input we use normal-component synoptic magnetograms from the US National Solar Observatory at Kitt Peak. These are used in two ways: to generate the initial condition for each run (a potential field extrapolation), and to determine the properties of newly-emerged bipoles (see \citet{yeates2007} for details). In this paper, we consider six simulation runs, each of $5.5$ Carrington rotations in duration. These periods are labelled A to F, and shown in Table \ref{tab:sims}. They are chosen based on the availability of the Kitt Peak data, so as to cover a range of phases of solar cycle 23 from one solar minimum to the next. The fourth column of Table \ref{tab:sims} shows the total number of bipoles emerged in each simulation run, while the right-most column indicates the instrument used for the magnetogram observations; the older vacuum telescope (KPVT) was replaced in 2003 by SOLIS (Synoptic Optical Long-term Investigations of the Sun).

\begin{table}
\caption{Simulation periods.}
\label{tab:sims}
\begin{tabular}{lllll}
\hline
Period & Carrington rotations & Dates & Bipoles & Instrument\\
\hline
{\bf A} (minimum) & 1911.5 to 1917 & 12-Jul-96 to 04-Jan-97 & 16 & KPVT\\
{\bf B} (rising phase) & 1948.5 to 1954 & 17-Apr-99 to 11-Oct-99 & 117 & KPVT\\
{\bf C} (maximum) & 1962.5 to 1968 & 03-May-00 to 27-Oct-00 & 122 & KPVT\\
{\bf D} (maximum) & 1974.5 to 1980 & 26-Mar-01 to 19-Sep-01 & 109 & KPVT\\
{\bf E} (declining phase) & 2018.5 to 2024 & 08-Jul-04 to 02-Jan-05 & 52 & SOLIS\\
{\bf F} (minimum) & 2067.5 to 2073 & 06-Mar-08 to 30-Aug-08 & 14 & SOLIS\\
\hline
\end{tabular}
\end{table}

\begin{figure} 
\centerline{\includegraphics[width=\textwidth]{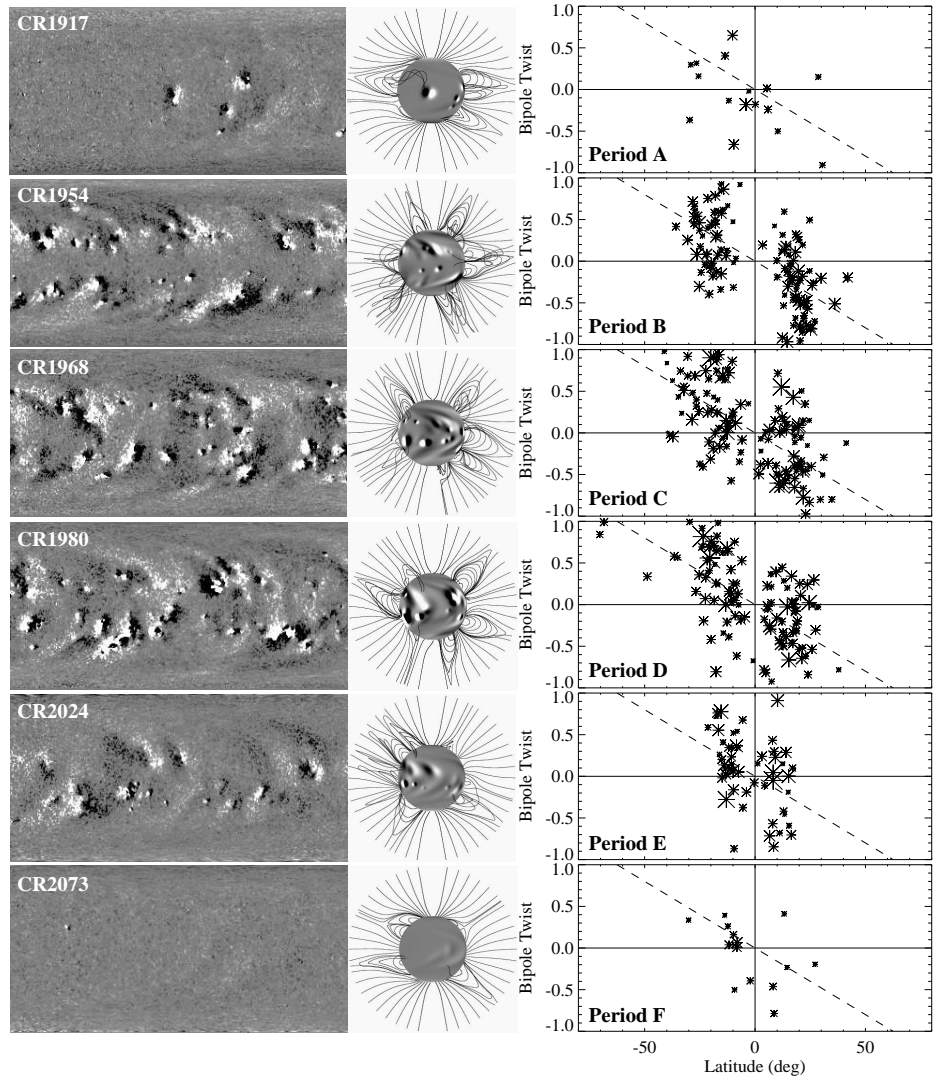}}
\caption{Summary of the six coronal magnetic field simulations. Left column shows example synoptic magnetograms from NSO/Kitt Peak in each period, middle column shows snapshots of the corresponding simulations, and right column shows helicity of emerging bipoles as a function of latitude. In the middle column, shading shows polarity of radial solar surface magnetic field, and lines show selected coronal field lines. In the right column, sizes of asterisks are proportional to the magnetic flux of each region.}\label{fig:simperiods}
\end{figure}

Figure \ref{fig:simperiods} shows example synoptic magnetograms from each of the six periods A to F (left column), along with snapshots of each simulation run (middle column). These snapshots are taken on the days corresponding to $180^\circ$ Carrington longitude in the synoptic magnetograms.

An important feature of our non-potential model is the ability to include non-zero magnetic helicity in emerging bipoles, through a parameter $\beta$ in the mathematical expression used to model them (\citet{yeates2008}, Equations (6)--(9)). The magnitude and, especially, the sign of this emerging bipole helicity have been shown to influence both the chirality of filament channels \citep{yeates2008,yeates2009a} and the ejection rate of magnetic flux ropes \citep{yeates2009b}. Unfortunately, the optimum value of the helicity parameter for each bipole is poorly constrained by the available observations, and must be arbitrarily chosen. In this paper, we depart from our previous practice of using the same value of $\beta$ for all bipoles in each hemisphere. Instead, we give each bipole a randomly chosen $\beta$ value. This is motivated by the limited existing observations of active region helicity, which show variation in magnitude and sign---both within individual active regions and between different regions \citep{hagino2004}---and also a negative gradient of helicity with latitude \citep{pevtsov1995}. To approximate these observations, we select the $\beta$ value for a bipole at latitude $\lambda^\circ$ from a normal distribution with mean $\beta_0=-0.4\lambda^\circ/25^\circ$ and standard deviation $0.4$. The selected values of $\beta$ for each bipole are shown as a function of latitude in the right column of Figure \ref{fig:simperiods}, for each simulation period. Here the dashed line shows the mean value $\beta_0(\lambda^\circ)$ of the normal distribution. Note that, in these plots, the sizes of the symbols indicate the relative amounts of magnetic flux in each bipole. Visible also is the sunspot ``butterfly diagram'' whereby emerging active regions progress toward the equator as the cycle progresses. While the random choice of helicity in each emerging bipole clearly limits the application of the model to the detailed structure of particular active regions, we are interested here in the global magnetic picture, and how this varies over the solar cycle in this model.

\section{Solar Cycle Variation}\label{sec:results}

\subsection{Global Magnetic Field Properties} \label{sec:field}

\subsubsection{Magnetic Flux}

\begin{figure} 
\centerline{\includegraphics[width=\textwidth]{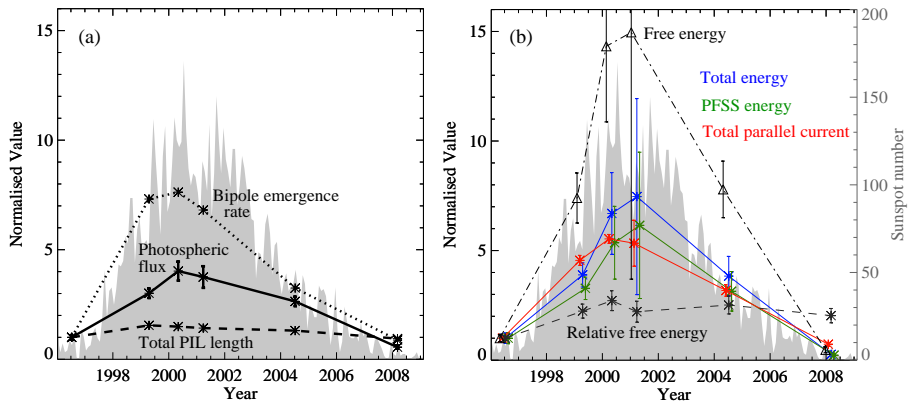}}
\caption{Cycle variation of simulated magnetic field properties: (a) properties of the photospheric magnetic field, and (b) properties of the 3D coronal magnetic field. Mean values of each quantity are taken over the last 100 days of each simulation period, and error bars give standard deviation between different days. Each quantity is normalized by its mean value for period A. Grey shading shows monthly smoothed sunspot number from SIDC.}\label{fig:cycleflux}
\end{figure}

It is clear from the magnetograms in Figure \ref{fig:simperiods} that there is more magnetic flux through the photosphere at solar maximum (periods C and D) than at solar minimum (periods A and F). In our simulations, the total (unsigned) radial magnetic flux through the photospheric boundary increases by a factor of 4 between minimum and maximum, from $1.4\times 10^{23}\,\textrm{Mx}$ in period A to $5.5\times 10^{23}\,\textrm{Mx}$ in period C. This relative increase is shown by the solid line in Figure \ref{fig:cycleflux}(a). Here we plot the mean value over the final 100 days of each simulation period, normalised by the value for period A.

\subsubsection{Complexity}

Not only is there more photospheric flux at maximum, but it is divided between more positive and negative regions, owing to the larger number of active regions emerging. In the simulations, the latter corresponds to the number of bipoles inserted, shown in the fourth column of Table \ref{tab:sims} and by the dotted line in Figure \ref{fig:cycleflux}(a). Again, the values are normalized by that for period A (16 bipoles). The rate of bipole emergence increases by a factor between 7 and 8 from minimum to maximum, similar to the observed sunspot number (shown by grey shading).

An alternative measure of the magnetic complexity is the total length of photospheric PILs at any one time in the simulation. This quantity increases only by a factor of about 1.5 from minimum to maximum, as shown by the dashed line in Figure \ref{fig:cycleflux}(a). Of course, the PIL length depends on the spatial resolution of the magnetic field. Here we consider PILs of the large-scale mean field \citep{vanballegooijen2000}, which remain well-defined entities over many days and are observed to have physical relevance as the locations above which filaments form.

\subsubsection{Non-Potentiality}

The deviation of our simulated magnetic field from potential may be characterised by integrating either magnetic energy or current over the 3D simulation volume. The total parallel current $|\mathbf{j}_0\cdot\mathbf{B}_0/B_0|$ (at a single time) increases by a factor of between 5 and 6 from minimum to maximum. The mean values for 100 days of each simulation period are shown by the red curve in red in Figure \ref{fig:cycleflux}(b). The additional coronal currents at solar maximum originate ultimately from the greater bipole emergence rate. Emerging bipoles inject current both directly, because they emerge twisted in our simulation, and indirectly, through their interaction with nearby regions and the further shearing of their fields by subsequent surface motions \citep{yeates2009b}.

The total magnetic energy $E_\textrm{np}$ is shown in blue in Figure \ref{fig:cycleflux}(b). As expected, it increases from minimum to maximum, from $8.9\times 10^{32}\,\textrm{ergs}$ in period A to $6.7\times 10^{33}\,\textrm{ergs}$ in period D---a factor of about 8, a similar relative increase to the bipole emergence rate. If we perform a sequence of potential-field source surface (PFSS) extrapolations from the simulated photospheric field, we may calculate the corresponding PFSS energy $E_\textrm{p}$ at each time (shown in green in Figure \ref{fig:cycleflux}b). The free energy of the simulated non-potential field may then be computed by subtracting $E_\textrm{p}$ from $E_\textrm{np}$. Shown by the triangles and dot-dashed line in Figure \ref{fig:cycleflux}(b), this free energy $E_\textrm{np}-E_\textrm{p}$ increases by a factor 15 from minimum to maximum, a greater relative increase than any other quantity. The actual mean values of free energy are $1.3\times 10^{32}\,\textrm{ergs}$ in period A and $2.0\times 10^{33}\,\textrm{ergs}$ in period D.

\begin{figure} 
\centerline{\includegraphics[width=\textwidth]{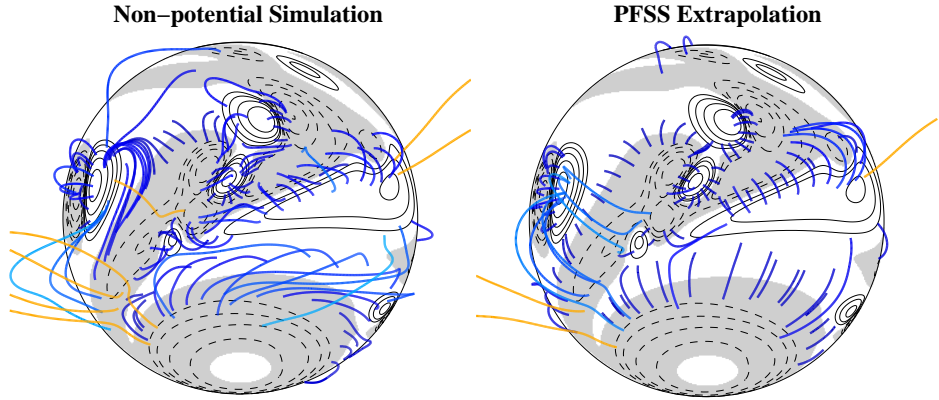}}
\caption{Contrast between non-potential simulation and potential field source-surface extrapolation at solar minimum (period A). The photospheric radial magnetic field is the same in both cases, shown by contours (white shading and solid lines for positive, gray shading and dashed lines for negative). Coloured lines show selected coronal magnetic field lines traced from the same photospheric starting points in each case; closed field lines are in blue (lighter shades for field lines reaching higher) and open field lines are in orange.}\label{fig:mincomp}
\end{figure}

Since the PFSS energy also increases over the cycle, a relative measure of the non-potentiality of the field is obtained by normalising the free energy by the PFSS energy. The resulting ``relative free energy'' $(E_\textrm{np}-E_\textrm{p})/E_\textrm{p}$ is shown by the dashed line in Figure \ref{fig:cycleflux}(b). Here the actual value for period A is $0.18$. Unlike the original free energy, the relative free energy does not show a clear modulation with solar activity. So although the total magnetic energy in the corona is much lower at minimum, the field in our simulation remains globally non-potential. To illustrate this point, Figure \ref{fig:mincomp} compares the magnetic field on the last day of period A, for the non-potential simulation (left) and a PFSS extrapolation from the same radial photospheric field (right). Here the Sun is viewed from the Southern hemisphere. Coloured lines show coronal magnetic field lines crossing various PILs, traced from the same locations on the photosphere in each field. It is clear that the field across most PILs is much more sheared in the non-potential simulation than in the PFSS extrapolation, even though the field strength is weak during this minimum period. This demonstrates how magnetic helicity is still generated in the corona at minimum, and not merely by emergence in the few remaining active regions. In the simulations, shearing of remnant fields by differential rotation generates helicity on a large scale, as evidenced by the south polar crown in Figure \ref{fig:mincomp}.

\subsection{Magnetic Flux Ropes} \label{sec:ropes}
Our method for identifying magnetic flux ropes in simulated magnetic field data is a slight modification of that described by \citet{yeates2009b}. As explained in that paper, we initially identify points on the computational grid satisfying certain criteria---so-called flux rope \emph{points}---before using a clustering algorithm to group these points into flux \emph{ropes}. 

To identify flux rope points, we compute the vertical magnetic tension force $T_r=(\mathbf{B}_0\cdot\nabla)B_{0r}/\mu_0$ and pressure gradient $P_r=-\partial_rB_0^2/(2\mu_0)$ at a subset of points in the computational domain between $r=R_\odot$ and $r=1.44R_\odot$, and select points by the following five criteria:
\begin{eqnarray}
P_r(r_{i-1},\theta_i,\phi_i) &<& -0.4 B_0^2(r_{i-1},\theta_i,\phi_i),\label{eqn:c1}\\
P_r(r_{i+1},\theta_i,\phi_i) &>& 0.4 B_0^2(r_{i+1},\theta_i,\phi_i),\label{eqn:c2}\\
T_r(r_{i-1},\theta_i,\phi_i) &>& 0.4 B_0^2(r_{i-1},\theta_i,\phi_i),\label{eqn:c3}\\
T_r(r_{i+1},\theta_i,\phi_i) &<& -0.4 B_0^2(r_{i+1},\theta_i,\phi_i),\label{eqn:c4}\\
|\mathbf{j}_0\cdot\mathbf{B}_0| &>& \alpha^*B_0^2,
\end{eqnarray}
where $\alpha^*=0.7\times10^{-8}\,\textrm{m}^{-1}$. The factor $B_0^2$ has been introduced in criteria (\ref{eqn:c1}) to (\ref{eqn:c4}) to make them independent of magnetic field strength, since the original criteria were developed for a period of high solar activity and found to miss flux ropes with very weak field at solar minimum. The new criteria are calibrated so as to reproduce the earlier flux rope statistics of \citet{yeates2009b}. That paper covers the same period as period B in this paper, but the simulation in this paper differs because the emerging bipoles have been given random twists, rather than all having the same value of $\beta$. This results in slightly fewer flux rope points overall and slightly fewer eruptions, due to the lower net helicity.

\begin{figure} 
\centerline{\includegraphics[width=\textwidth]{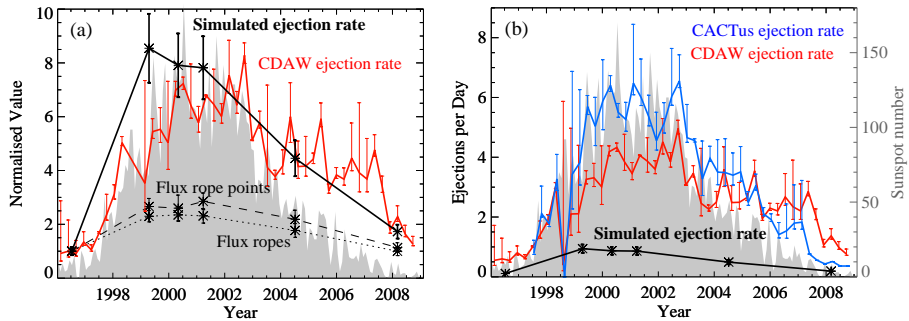}}
\caption{Cycle variation of simulated magnetic flux ropes and observed CMEs: (a) normalized by values for period A, and (b) actual ejection rates per day. Values from simulations are averaged over the last 100 days of each period (error bars give the standard deviation between different days). Observed CME rates from the CDAW and CACTus catalogues are shown averaged over 100-day periods. Estimation of their errors is described in the text. Grey shading shows the monthly smoothed sunspot number.}\label{fig:cycleropes}
\end{figure}

Comparing the six simulations in this paper, the number of flux rope points is found to vary over the solar cycle. The mean number of flux rope points per day for each simulation period is shown in Figure \ref{fig:cycleropes}(a) by the dashed line, normalized by the mean value for period A (692 points). It increases by a factor of about 2.5 from minimum to maximum. The dotted line in Figure \ref{fig:cycleropes}(a) shows the number of flux \emph{ropes} present at any one time, determined on each day by clustering the flux rope points \citep[as described][]{yeates2009b}. Normalised by the mean value for period A (19.2 ropes), this shows a similar factor-two increase. 

\begin{figure} 
\centerline{\includegraphics[width=0.7\textwidth]{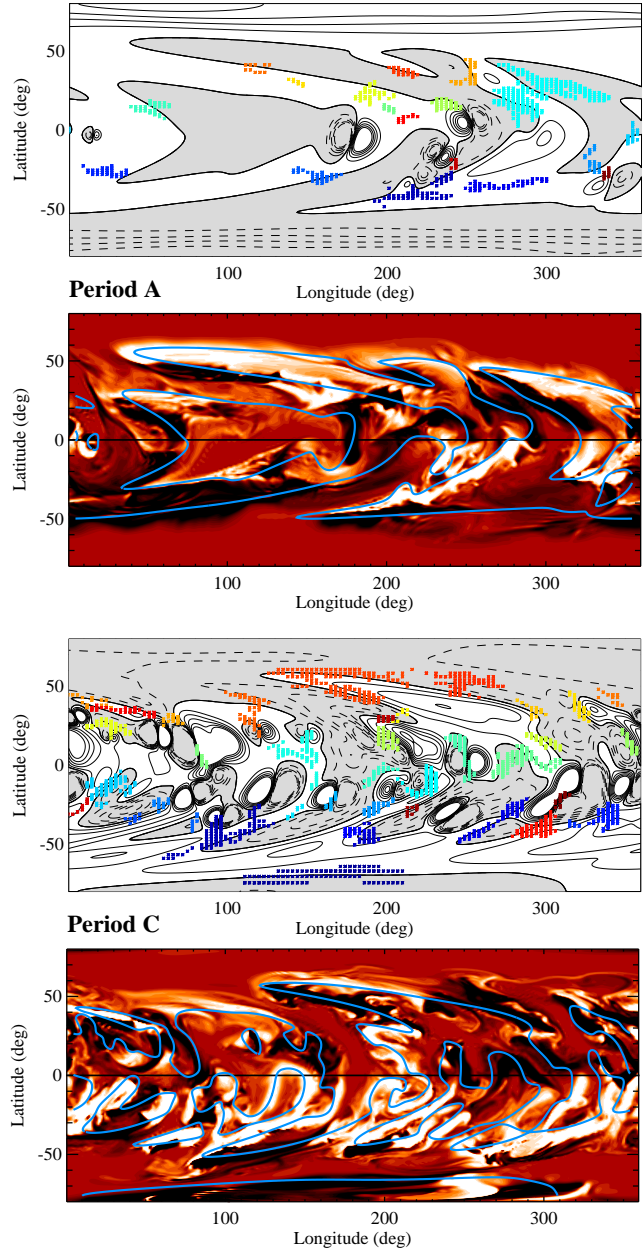}}
\caption{Comparison of low-coronal magnetic field on final day of periods A and C. The top panels for each period show the photospheric radial magnetic field (solid contours/white for positive, dashed contours/gray for negative) and locations of flux ropes on this day (coloured points). The bottom, red shaded, panels show the distributions of current helicity $\alpha$ at height $12\,\textrm{Mm}$ (white indicates positive, black negative, saturation level $\pm 20\times10^{-8}\,\textrm{m}^{-1}$), with photospheric PILs in blue. }\label{fig:comp}
\end{figure}

This doubling of the number of flux ropes present at any one time may be understood from the increase in total photospheric PIL length (which was by a factor of about 1.5). This is illustrated by Figure \ref{fig:comp}, which compares the magnetic structure on the final day of periods A (top two panels) and C (bottom two panels), representing minimum and maximum respectively. The upper panel in each pair shows flux rope points overlayed on the photospheric magnetic field (with differently coloured clusters corresponding to different flux ropes), while the lower panels show the distribution of current helicity (density) $\alpha=\mathbf{j}_0\cdot\mathbf{B}_0/B_0^2$ over the spherical surface at height $12\,\textrm{Mm}$, in the low corona. Flux ropes are seen to form at locations where current helicity is concentrated above photospheric PILs, which is a key feature of this model. It is therefore natural that at maximum---when there are more PILs---there should be more flux ropes. Indeed, both the total PIL length and the integral of $|\alpha|$ over this spherical surface show increases of about a factor 1.5 from minimum to maximum, comparable to the number of flux ropes.

We remark that the precise origin of the magnetic helicity that becomes concentrated above PILs is still an open question, and need not be the same for PILs everywhere on the Sun. \citet{yeates2009a} found that sheared magnetic fields overlying PILs in this model arise purely from differential rotation when the PILs are at high latitudes, but over low-latitude PILs shear may also arise from the emergence of twisted magnetic fields in active regions, and from the interaction between regions of strong magnetic field.

\begin{figure} 
\centerline{\includegraphics[width=0.8\textwidth]{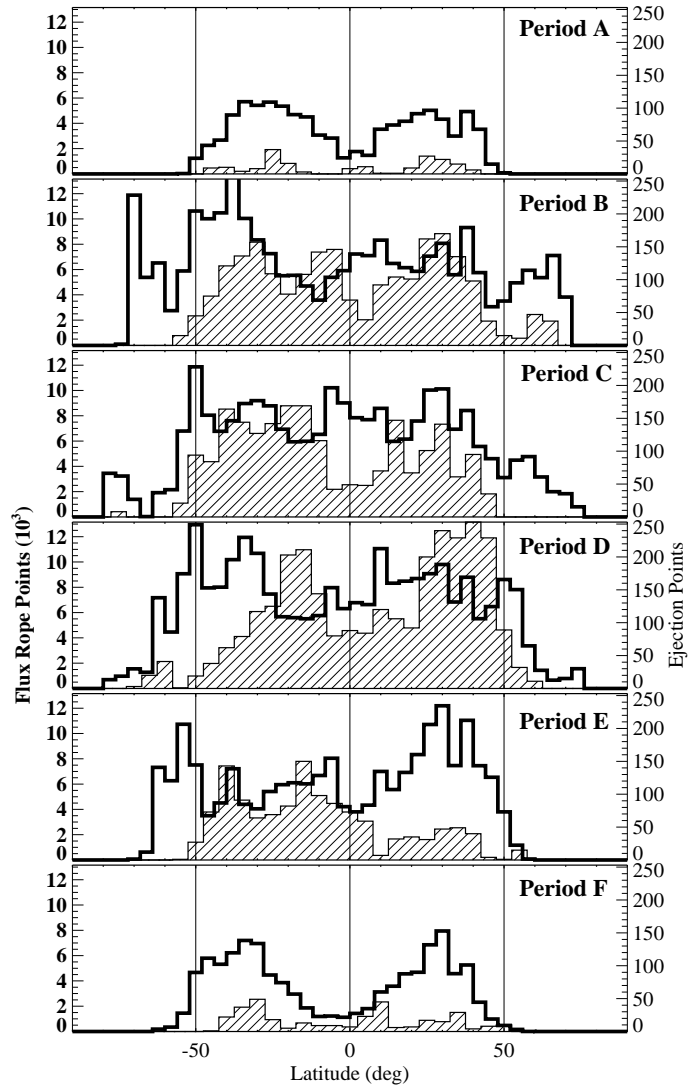}}
\caption{Histograms showing latitude distribution of flux rope points (thick lines) and ejection points (filled), for final 100 days of each simulation period. Bin sizes are $4^\circ$ and $5^\circ$ respectively.}\label{fig:ropelats}
\end{figure}

A consequence of the dependence of flux rope locations on PILs is that the latitude distribution of flux rope points varies over the solar cycle. This is shown by the thick histograms in Figure \ref{fig:ropelats}. At minimum (periods A and F) there is a bimodal distribution at active latitudes, whereas in periods B to E the distribution of flux ropes extends both across the equator and to higher latitudes. High latitude (above $60^\circ$) flux ropes are particularly prevalent in period B. This is due to the poleward migration of the polar crowns, prior to the polar field reversal.

\subsection{Flux Rope Ejections} \label{sec:ejections}

To detect flux rope ejections in each simulation run, we have again used the procedure described by \citet{yeates2009b}, selecting those flux rope points with $v_{0r}>0.5\,\textrm{km}\,\textrm{s}^{-1}$ in the magneto-frictional code, and clustering in both space and time into separate ejection events. With the additional factor $B_0^2$ in the detection of flux rope points (Section \ref{sec:ropes}), we found that performance of the ejection detecting algorithm was optimized if the parameters were slightly adjusted: we now require that clusters are separated by at least 5 days in time, and have at least 8 points. 

To look at how the simulated ejection rate varies over the solar cycle, we count the number of ejections in the last 100 days of each simulation period. This varies from a minimum of 11 in period A up to a maximum of 94 in period B. Note that this ejection rate for period B is lower than those in the simulation runs of \citet{yeates2009b} which had uniform signs of bipole twist of magnitude $|\beta|=0.2$ or $0.4$ in each hemisphere. In the present simulation, nearby bipoles with opposite signs of $\beta$ tend to mutually ``cancel'' some of their helicity. Figures \ref{fig:cycleropes}(a) and (b) show how this simulated ejection rate varies over the solar cycle, respectively normalised (by the value for period A) and un-normalised. In this case the error bars show estimated errors in the number of ejections resulting from the automated detection procedure \citep[see][]{yeates2009b}.

This eightfold increase in the rate of ejections is similar to that of the total magnetic energy in the corona, or that of the bipole emergence rate. We now make two comparisons. Firstly, the cycle increase in ejection rate is larger than that in the number of flux rope points present at any one time (Section \ref{sec:ropes}). Therefore a higher proportion of flux ropes lose equilibrium in a given interval at maximum than at minimum, so flux ropes at maximum have shorter lifetimes on average.
Secondly, the increase in ejection rate is only about half that in free magnetic energy (Section \ref{sec:field}). Hence the total free energy, as plotted in Figure \ref{fig:cycleflux}(b), does not determine directly the number of flux rope ejections occurring, but rather the local structure and evolution of the magnetic field must be taken into account. In particular, the similar increases in ejection and bipole emergence rates suggest that newly-emerging active regions are important.

We can see that the increased rate of ejections is somehow related to newly-emerged regions by inspecting the latitude distribution of ejection points (the filled histograms in Figure \ref{fig:ropelats}). In Periods B to E, the distribution of ejection points is more bimodal than that of flux rope points, and has fewer points at higher latitudes, with only a few ejections on the polar crowns. Indeed, a similar bimodal latitude distribution is found in observations of disappearing filaments \citep{pojoga2003}, of prominence eruptions \citep{gopalswamy2003}, and of EUV source regions associated with CMEs \citep{plunkett2001,cremades2004}.

There are several mechanisms by which increased flux emergence causes locally increased propensity for flux rope ejection. Firstly, increased emergence injects a greater amount of magnetic energy and helicity, leading to more rapid growth of flux ropes and hence earlier loss of equilibrium. This is consistent with the finding of similar relative increases in both ejection rate and total magnetic energy. Secondly, active regions have relatively more magnetic energy than the overlying field, so flux ropes in or around active regions are more likely to lose equilibrium. Finally, there is the possiblity of destabilisation of existing flux ropes by the nearby emergence of new active regions.

\section{Observed CME Rate} \label{sec:obs}

How does the rate of flux rope ejections produced by the model compare to the observed CME rate? First, note that care should be taken with observed CME rates, because these depend both on instrument sensitivity \citep{cremades2007} and on the processing of the data, including the definition of a CME \citep[{\it e.g.},][]{yashiro2008}.

These limitations notwithstanding, we have determined two observed rates over cycle 23, based on the published CDAW and CACTus catalogues, both using data from LASCO \citep[Large Angle and Spectrometric Coronagraph;][]{brueckner1995}. The CDAW (Coordinated Data Analysis Workshops) catalogue \citep[][available at \url{http://cdaw.gsfc.nasa.gov/CME list/}]{yashiro2004} is the standard manually-compiled list of LASCO CMEs; we filter out CMEs with apparent width less than $15^\circ$ or greater than $270^\circ$. The resulting number of CMEs per day is shown by the red curves in Figures \ref{fig:cycleropes}(a) and (b). The blue curve in Figure \ref{fig:cycleropes}(b) shows the rate from the alternative CACTus catalogue (\url{http://sidc.be/cactus}), which is based on the same LASCO observations but uses an automated technique to detect radial motion in height-time maps \citep{robbrecht2004}. We omit so-called ``marginal cases'' in the CACTus data (see \url{http://sidc.oma.be/cactus/scan/}). The error bars on these observed rates in Figure \ref{fig:cycleropes} are determined taking into account instrument data gaps, following the method of \citet{stcyr2000}. Namely, the lower bar is the actual observed number of CMEs, the data point itself has additional CMEs at this same rate to fill any data gaps, and the upper bar fills the data gaps with CMEs at the maximum rate observed over any 24-hour period. 

Figure \ref{fig:cycleropes}(a) shows an appealing agreement between the relative increases of simulated and observed (CDAW) ejection rates over the cycle. However, Figure \ref{fig:cycleropes}(b) shows that the actual simulated rate is much lower than the observed rates from either catalogue (which agree quite well). The ratio of simulated to observed ejection rates remains roughly constant, at about $0.2$, over the cycle. The implications of this shortfall for possible mechanisms of CME initiation are considered elsewhere \citep{yeates2009}. Note that this ratio of 0.2 is lower than that found by \citet{yeates2009b} in their optimum case; partly because they omitted ``poor events'' in the CDAW catalogue, and partly because the bipoles in this paper have been given random twists, rather than all having the same value of $\beta$ as in the earlier simulations. This results in lower net helicity, which was shown by \citet{yeates2009b} to result in fewer eruptions.

\section{Conclusions}\label{sec:conclusions}

As a step toward removing the potential-field assumption in modelling the global coronal magnetic field, we have investigated a quasi-static model allowing for electric currents. This model is driven by photospheric magnetic observations and follows the build-up of magnetic helicity as magnetic flux emerges and is transported by surface motions. In this paper, we simulate six distinct periods during solar cycle 23, so as to study how the magnetic field structure predicted by this model varies over the solar cycle.

A key feature of this model is the transport of magnetic helicity and its concentration into twisted magnetic flux ropes above photospheric PILs. The solar cycle variation of these flux ropes in the model may be characterised as follows:
\begin{enumerate}
\item The number of flux ropes present at any one time doubles between minimum and maximum of the solar cycle. This follows the total length of photospheric PILs.
\item The rate of flux rope ejections---or losses of equilibrium---increases by a factor of eight between minimum and maximum. This is lower than the relative increase in free magnetic energy, but similar to the relative increases in total magnetic energy and total parallel current, both of which follow the number of emerging active regions.
\end{enumerate}
The difference between the relative increases in ejection rate and in the number of flux ropes implies that, at maximum, the flux ropes have shorter lifetimes before losing equilibrium. This is a consequence of the higher rate of active region emergence.

Although there are fewer flux rope ejections at minimum, it is not the case that the corona in our model is everywhere close to potential. Even in the absence of many emerging bipoles, shearing of pre-existing coronal field by differential rotation generates current on a large scale at all latitudes. A weak background field originating in earlier decayed active regions is present even in the recent extended minimum (modelled by our period F in 2008). Our model assumes that differential rotation is equally effective at shearing this weaker field, so that, for example, we find systematic sheared arcades on the high-latitude polar crowns. Interestingly, although sheared arcades are formed, the low reconnecting flux over these PILs in 2008 prevents detached helical flux ropes from forming at many locations, unlike at solar maximum. The true nature of the magnetic field above the polar crown PILs remains uncertain. Our simulations predict that differential rotation will develop positive and negative helicity over such east-west PILs in the northern and southern hemispheres respectively. However, the (few) existing observations of the magnetic field in polar crown prominences imply the opposite sign of helicity in each hemisphere \citep{leroy1983}. In our quasi-static model, the correct sign of helicity is only obtained at such PIL locations if additional emergence of axial flux is included \citep{vanballegooijen2000}. This was not included in the present paper because it lacks observational justification. A concerted effort to obtain further observations of the magnetic field structure at high latitudes is needed in order to resolve this issue and hence test whether additional sources of helicity are required in addition to those in our existing model.

An important question that may be addressed by a time-dependent non-potential model of the coronal magnetic field is the initiation of CMEs. Our model allows us to test whether the loss of equilibrium of flux ropes formed by quasi-static shearing and flux cancellation is sufficient to account for the observed CME rate. For the present form of the model the answer is negative: the ratio of simulated to observed ejection rate (from CDAW) remains about $0.2$ over the whole solar cycle. Even in our earlier (less realistic) simulations where all emerging bipoles were given the same sign of helicity in each hemisphere, the ratio increases only to about $0.25$. This suggests strongly that the formation of flux ropes by quasi-static shearing of the large-scale magnetic field is incapable of initiating all CMEs. We should point out, however, that the CDAW rate for period F (in 2008) may be too high (as discussed in Section \ref{sec:ejections}), in which case the model may account for a greater proportion of CMEs in the recent very quiet period. Nevertheless, it is clear that the present model lacks the detail to adequately describe initiation of all CMEs.

More detailed comparison with observed low-coronal CME source regions \citep[described by][]{yeates2009} reveals that many of the additional observed CMEs are located in active regions, with individual active regions frequently producing multiple CMEs in the same day. Our global quasi-static simulations cannot produce such dynamic events with the present form of input data (updated once per Carrington rotation). The quasi-static approach could be developed in future to incorporate driving from photospheric magnetograms on much shorter spatial and time scales, in order to test whether the energetic restructuring of the corona following flux emergence may be adequately described by a quasi-static model, or whether fully dynamic MHD modelling is required. From the global perspective, this more detailed modelling of active regions should allow their helicity content, and hence the global transport of helicity, to be better constrained by observations. This has important implications both for models of the coronal magnetic structure and, ultimately, for the origin of the Sun's magnetic field in the solar dynamo \citep{seehafer2003}.

\begin{acks}
We thank A.A. van Ballegooijen, D.H. Mackay, A.N. Wilmot-Smith, D.I. Pontin and an anonymous referee for useful suggestions, and D.H. Mackay also for the use of parallel computing facilities obtained through a Royal Society research grant. ARY acknowledges financial support from NASA contract NNM07AB07C at SAO, and from the UK STFC at Dundee. The visit of JAC to SAO was supported by NASA grant NNX08AW53 and by NSF grant ATM-0851866 for ``REU site: Solar Physics at the Harvard-Smithsonian Center for Astrophysics.'' Magnetogram data from NSO/Kitt Peak were produced cooperatively by NSF/NOAO, NASA/GSFC and NOAA/SEL. SOLIS data used here are produced cooperatively by NSF/NSO and NASA/LWS.
\end{acks}

\end{article} 
\end{document}